\documentclass[sigconf,screen,authorversion]{acmart}

\usepackage{listings}
\usepackage{pifont}

\AtBeginDocument{%
  }

\copyrightyear{2026}
\acmYear{2026}
\setcopyright{cc}
\setcctype{by}
\acmConference[ECOOP-C '26]{ECOOP-Companion}{June 29--30, 2026}{Brussels, Belgium}
\acmBooktitle{ECOOP-Companion (ECOOP-C '26), June 29--30, 2026, Brussels, Belgium}
\acmDOI{10.1145/3849733.3849776}
\acmISBN{979-8-4007-2540-1/2026/06}

\definecolor{OliveGreen}{cmyk}{0.64,0,0.95,0.40}
\definecolor{Brown}{cmyk}{0,0.81,1,0.60}

\lstdefinelanguage[]{Emfrp}{
  sensitive = true,
  morekeywords = {
    module, material, in, out, use, init, 
    node, newnode, data, type, func, where,
    if, then, else, let,
    case, of,
    record
  },
  morecomment=[l]{\#},
  otherkeywords = {
    @last
  },
  escapechar=?
}

\lstdefinestyle{default}{%
  basicstyle={\ttfamily},
  keywordstyle={\color{Brown}\bfseries},
  commentstyle={\color{OliveGreen}\small},
  stringstyle={\color{blue}},
  emphstyle={\bfseries},
  tabsize=2,
  keepspaces=true,
  columns=fixed,
  xrightmargin=0.5em,
  xleftmargin=1em,
  breaklines=true,
  basewidth={0.45em},
  lineskip=-0.4ex,
  captionpos=b,
  frame=single,
  numbers=left,
  numbersep=5pt,
  numberstyle=\tiny
}
\lstdefinestyle{small}{
  style=default,
  basicstyle={\ttfamily\small},
  basewidth={0.5em}
}
\begin{document}

\title[Multi-Mode Debugging for FRP-Based Embedded Systems]{Multi-Mode Debugging\\for FRP-Based Embedded Systems}

\author{Yugo Otani}
\orcid{0009-0003-8545-7180}
\affiliation{
  \department{Department of Computer Science}
  \institution{Institute of Science Tokyo}
  \city{Tokyo}
  \country{Japan}}
\email{yugo@psg.comp.isct.ac.jp}
\author{Sosuke Moriguchi}
\orcid{0000-0002-4153-4514}
\affiliation{
  \department{Department of Computer Science}
  \institution{Institute of Science Tokyo}
  \city{Tokyo}
  \country{Japan}}
\email{chiguri@comp.isct.ac.jp}
\author{Takuo Watanabe}
\orcid{0000-0001-7470-3428}
\affiliation{
  \department{Department of Computer Science}
  \institution{Institute of Science Tokyo}
  \city{Tokyo}
  \country{Japan}}
\email{takuo@comp.isct.ac.jp}


\begin{abstract}
Emfrp is a functional reactive programming (FRP) language designed for small-scale embedded systems. Time-varying values are the primary abstraction mechanism in FRP and enable concise descriptions of reactive behavior. In practice, however, Emfrp programs are compiled into C and combined with platform-specific input/output components written in C or C++. Consequently, developers must debug the resulting mixed C/C++ program using conventional debuggers such as GDB, even though the application logic is written in Emfrp. This situation creates an abstraction gap between the source-level FRP program and the executable system.

This paper presents a multi-mode debugging framework for Emfrp-based embedded applications. The framework supports debugging at the level of Emfrp abstractions while also allowing inspection of platform-specific C/C++ I/O code. Our approach uses a source mapping technique that relates Emfrp constructs to corresponding locations in the compiled program.
A case study on an ESP32 microcontroller using representative debugging scenarios demonstrates improved debugging efficiency.
\end{abstract}




\maketitle

\section{Introduction}
\label{sec:Introduction}

Debugging reactive systems written in functional reactive programming (FRP) can be challenging, particularly in embedded environments. FRP programs describe system behavior as relationships among \emph{time-varying values} (also known as \emph{signals})~\cite{Elliott:1997aa,Bainomugisha:2013aa}, that is, values that evolve over time and can conceptually be viewed as functions of time. By representing reactive behavior as data dependencies among such values rather than explicit event-driven control flow, FRP enables developers to describe reactive systems at a higher level of abstraction.

However, this abstraction also introduces difficulties during debugging. Errors in FRP programs are often embedded in temporal dependencies among reactive values, making it difficult to determine which update of which value at what time caused an unexpected behavior. This challenge becomes even more pronounced in embedded environments, where runtime observation and instrumentation are often limited.

Emfrp~\cite{Sawada:2016aa} is an FRP language designed for small-scale embedded systems. The language provides built-in support for time-varying values and allows their current values to be referenced directly in expressions. 
To support resource-constrained platforms, Emfrp prohibits recursive functions and recursive data structures, enabling static determination of memory usage.

The Emfrp compiler translates Emfrp programs into C source code. The generated code implements the FRP runtime that periodically updates time-varying values according to their definitions. Platform-specific input/output functionality is implemented separately by developers in C or C++. The generated C code and the developer-written I/O code are then compiled together using a C compiler for the target platform to produce an executable for the embedded device.

Currently, debugging Emfrp programs relies on conventional embedded debugging techniques such as JTAG-based debugging with tools like GDB. In this setting, developers must debug the generated C program directly, forcing them to manually relate high-level Emfrp constructs to low-level C code. This mismatch between abstraction levels creates a significant obstacle to effective debugging.

To address this problem, we propose a multi-mode debugging framework for Emfrp-based embedded applications. In this framework, a mode corresponds to the level of abstraction at which debugging is performed. The framework supports two modes: debugging at the level of Emfrp source code and debugging at the level of platform-specific C/C++ I/O code. By bridging these levels, developers can investigate system behavior in terms of Emfrp abstractions while still accessing the underlying implementation when necessary. Based on this framework, we implement Emdb, a debugger that provides Emfrp-aware debugging operations and integrates them with an existing C debugger.

The contributions of this paper are as follows:
\begin{itemize}
\item FRP-aware debugging operations tailored to Emfrp programs.
\item Source mapping and command translation connecting Emfrp code with generated C code and existing C debuggers.
\item Evaluation on an ESP32 platform demonstrating improved debugging efficiency.
\end{itemize}

\section{Motivation}\label{sec:Mtivation}

\subsection{Emfrp}
Emfrp~\cite{Sawada:2016aa} is a statically typed, purely functional reactive programming (FRP) language designed for small-scale embedded systems, typically targeting microcontrollers. The language is designed so that programs can run even in highly resource-constrained environments with only a few kilobytes of RAM
In typical deployments, target platforms provide from a few kilobytes to several tens of kilobytes of RAM and up to a few megabytes of flash memory. To support such environments, Emfrp programs are designed to have a small memory footprint and to allow their runtime memory usage to be determined statically. The Emfrp compiler\footnote{\url{https://github.com/psg-titech/emfrp}} translates Emfrp source code into C code, which is compiled together with platform-specific input/output code written in C or C++ using a C compiler for the target platform to produce an executable for the embedded device.

In Emfrp, time-varying values are represented as \emph{nodes}, which form the basic abstraction of FRP computation. Each node represents the current value of a time-varying quantity that evolves over time. 
Nodes representing inputs from the external environment are called \emph{input nodes}; nodes representing outputs to external devices are called \emph{output nodes}; and all other nodes are called \emph{intermediate nodes}.
An Emfrp program is organized as a module consisting of a header that declares input and output nodes followed by definitions of types, functions, or nodes. Nodes are defined using the following syntax:
\begin{center}
\lstinline[language=Emfrp,mathescape=true]|node init[$c$] $\mathit{name}$ = $e$|
\end{center}
Here, $\mathit{name}$ denotes the node being defined, the optional \lstinline[language=Emfrp,mathescape=true]|init[$c$]| clause specifies the initial value $c$ of the node, and $e$ is an expression that computes the current value of the node.


\begin{lstlisting}[style=small,language=Emfrp,float,caption={DoubleClick Module},label={code:doubleclick}]
module DoubleClick
   in btn(False) : Bool,  # button
      nowMs: Int          # msec from reset
   out led : Bool

node init[False] led =
  if isDouble then !led@last else led@last

type Opt[a] = Some(a) | None  # option type

# Time when the button press started
node init[None] pressTime =
   if !inCooldown && !btn@last && btn
   then Some(nowMs)
   else pressTime@last

# Whether a click is detected
# (press duration <= 250 ms)
node clicked = {
   up = btn@last && !btn
   pressTime@last of:
      Some(t) -> 
        !inCooldown && up && nowMs - t <= 250
      None -> False
}

# Time when a click was detected
node init[None] clickTime =
  if clicked then Some(nowMs)
             else clickTime@last

# Whether a double click is detected
# (second press within 500 ms after a click)
node init[False] isDouble : Bool = {
   dwn = !btn@last && btn
   clickTime@last of:
      Some(t) -> dwn && nowMs - t <= 500
      None -> False
}

# Time when the cooldown period ends
# after a double click is detected
node init[None] cooldownUntil = 
   if isDouble then Some(nowMs + 500) 
               else cooldownUntil@last

# Whether the system is currently in the
# cooldown period
node inCooldown =  cooldownUntil@last of:
   None -> False
   Some(t) -> nowMs < t
\end{lstlisting}










Listing~\ref{code:doubleclick} shows an example Emfrp program that toggles an LED each time a button is double-clicked. To avoid unintended repeated activations, the program ignores button inputs for a short cooldown period after a double-click is detected. In this module, the node \texttt{btn}, which represents the current button state, and the node \texttt{nowMs}, which represents the elapsed time since system startup in milliseconds, are declared as input nodes. The node \texttt{led}, representing the LED state, is declared as an output node. The remainder of the program defines intermediate nodes that detect button presses, determine whether a click or double-click has occurred, and maintain the cooldown state.

Emfrp provides the keyword \lstinline[language=Emfrp]|@last| to access the \emph{previous value} (conceptually, the value from an infinitesimal amount of time ago) of a node.
For example, the expression \lstinline[language=Emfrp]|!btn@last && btn| becomes true only when the button transitions from released to pressed, corresponding to the rising edge of the button signal. Nodes can also maintain state using this mechanism. For instance, the definition of \texttt{led} toggles the LED state whenever a double-click is detected while preserving the previous state otherwise.

\begin{figure}
\centering\includegraphics[width=\linewidth]{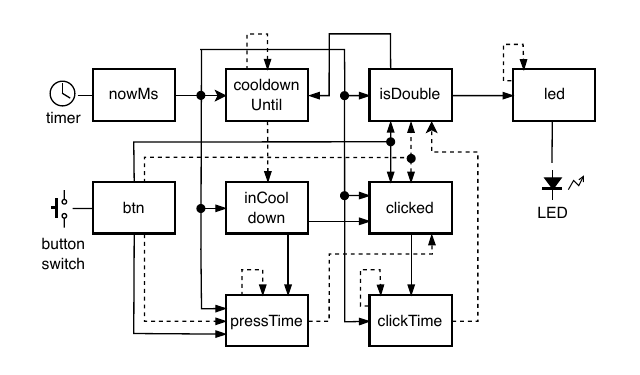}
\caption[Dependency Graph of the Nodes in DoubleClick]{Dependency Graph of the Nodes in \texttt{DoubleClick}}\label{fig:DependencyGraph}
\Description{Dependency graph of the DoubleClick module. Nodes represent time-varying values, solid arrows represent direct dependencies, and dashed arrows represent dependencies via \texttt{@last}.}
\end{figure}

Figure~\ref{fig:DependencyGraph} shows the dependency graph of the nodes in the program of Listing~\ref{code:doubleclick}. In this graph, each vertex represents a node (i.e., a time-varying value) in the Emfrp program, and each directed edge represents a dependency between nodes. A solid edge indicates a dependency through a direct reference, while a dashed edge represents a dependency through a reference using \lstinline[language=Emfrp]|@last|. The graph formed by solid edges must be acyclic; that is, the direct dependency graph must be a directed acyclic graph (DAG). The Emfrp compiler performs a topological sort of this DAG to obtain a linear ordering of nodes and generates code that updates node values in that order. In Emfrp, one complete update of this ordered list of nodes is called an iteration. 

\subsection{Debugging Emfrp Programs}
In practice, debugging Emfrp programs follows the standard workflow used for embedded systems. The development PC is connected to the target microcontroller through a debugging interface such as JTAG, and a debugger running on the host PC (e.g., GDB) interacts with the device through a debug server such as OpenOCD\footnote{\url{https://openocd.org}}. As described in Section~\ref{sec:Introduction}, however, the program being debugged is not the original Emfrp program but the C code generated by the Emfrp compiler together with platform-specific C/C++ I/O code. As a result, developers must analyze a mixed C/C++ program whose structure differs significantly from the original Emfrp source.

\begin{lstlisting}[style=small,language=C,float,caption={Compiled Code of the Node \texttt{clicked}},label={code:compiledcode}]
static int node_clicked(int v_inCooldown,
    int v_nowMs,
    struct Opt_Int v_pressTime_at_last, 
    int v_btn_at_last, int v_btn, int* output) {
  int _tmp007;
  int _tmp006;
  _tmp007 = _anpersand_anpersand(v_btn_at_last,
                _at__exclamation_(v_btn));
  if (1) {
    int pvar3_up = _tmp007;
    int _tmp008;
    if (v_pressTime_at_last.tvalue_id == 0) {
      int pvar4_t = 
          v_pressTime_at_last.value.Some.member0;
      _tmp008 = _anpersand_anpersand(
        _at__exclamation_(v_inCooldown),
        _anpersand_anpersand(pvar3_up, 
        _lt__eq_(_minus_(v_nowMs, pvar4_t),
                 250)));
    }
    else {
      _tmp008 = 0;
    }
    _tmp006 = _tmp008;
  }
  *output = _tmp006;
  return 1;
}
\end{lstlisting}

Listing~\ref{code:compiledcode} shows a fragment of the C code generated by the current Emfrp compiler that corresponds to the node \texttt{clicked} in Listing~\ref{code:doubleclick}. 
Although the generated code faithfully implements the semantics of the Emfrp program, its structure differs considerably from the original source. For example, the compiler introduces temporary variables (such as \verb|_tmp007|) and additional control structures for code generation purposes. These artifacts do not appear in the original Emfrp program and make it difficult for developers to identify which conditions determine the value of each node.

Another challenge arises from differences in identifier naming and data representation. Although the compiler attempts to preserve some correspondence between identifiers in Emfrp and those in the generated C code, the names are not always identical and may vary depending on the context in which they are used. Furthermore, user-defined algebraic data types in Emfrp are translated into combinations of C integers, structures, and enumerations. Developers must therefore reconstruct the original Emfrp-level data structures by examining how these low-level representations are accessed and manipulated in the generated C code.

A further complication stems from the execution model of Emfrp programs. As described in the previous subsection, nodes are evaluated in a topologically sorted order of their dependency graph. The keyword \lstinline[language=Emfrp]|@last| allows a node to refer to its value from the previous iteration. At runtime, the system maintains both the current value and the previous value of each node. However, this distinction is not always explicit in the generated C code, making it difficult for developers to understand whether a variable corresponds to a current value or a previous one when debugging at the C level.

These factors significantly increase the cognitive burden of debugging Emfrp programs at the C level. Developers must simultaneously reason about the original Emfrp program, the generated C code, and the runtime representation of program state. Consequently, much of the debugging effort is spent reconstructing the correspondence between abstraction levels rather than diagnosing the actual cause of the bug. This situation motivates the need for a debugging framework that allows developers to observe and control program execution directly at the Emfrp level.

\section{Bug Analysis and Requirements}

\subsection{Taxonomy of Bugs in Emfrp Programs}
To design an effective debugging framework for Emfrp, it is useful to classify the kinds of bugs that arise in Emfrp-based systems and to identify what forms of observation and control are needed to diagnose them. An Emfrp application can be viewed as consisting of three layers: the \emph{FRP layer}, which is written in Emfrp and implements the main application logic; the \emph{I/O layer}, which is written in C and handles interaction with external devices; and an \emph{intermediate layer}, which contains primitive data types and functions implemented in C. Because these layers differ in both language and role, the kinds of bugs that arise in them, and the debugging support required to diagnose those bugs, also differ. In this paper, we focus primarily on the FRP layer and the I/O layer.

At the FRP layer, bugs can be divided into two broad categories: \emph{time-independent bugs} and \emph{time-dependent bugs}. Time-independent bugs are those that do not essentially depend on references to previous values through \lstinline[language=Emfrp]|@last|. Typical examples include errors in conditional branches, logical expressions, arithmetic expressions, or argument order. Conceptually, these are similar to ordinary bugs in functional programs: the value computed in a single evaluation step is incorrect. To localize such bugs, a debugger must reveal which branch was taken, the values of local variables, and the values of relevant subexpressions at runtime.

Time-dependent bugs arise from the temporal structure of Emfrp programs with \lstinline[language=Emfrp]|@last|. Without \lstinline[language=Emfrp]|@last|, an Emfrp module behaves essentially like a combinatorial logic circuit: its outputs are functions of the current inputs and it cannot express stateful behavior over time. In practice, \lstinline[language=Emfrp]|@last| is used for at least three purposes: implementing state, detecting moments of change, and breaking dependency cycles. In Listing~\ref{code:doubleclick}, nodes \texttt{cooldownUntil}, \texttt{pressTime}, \texttt{clickTime}, and \texttt{led} use \lstinline[language=Emfrp]|@last| to retain state across updates; these self-dependencies appear as dashed self-loops in Figure~\ref{fig:DependencyGraph}. The expression \lstinline[language=Emfrp]|!btn@last && btn| (lines 13 and 35 in Listing~\ref{code:doubleclick}) illustrates the second use, namely detecting the instant at which the button is pressed. Other uses of \lstinline[language=Emfrp]|@last| in the example correspond to resolving cyclic dependencies. In these cases, typical bugs include missing or incorrect conditions, inconsistent initial values, and mistakes in where \lstinline[language=Emfrp]|@last| is applied. Such bugs are often harder to diagnose because the relevant cause may lie not only in the current expression but also in the relationship between current and previous values.

The I/O layer has a different character. It is written in C and is responsible for interaction with external devices. Although arbitrary C code can be used in this layer, Emfrp assumes that developers do not introduce long blocking operations that prevent the progress of iterations or asynchronous updates that interfere with the FRP layer, for example through interrupts that destructively modify shared state. Typical bugs in this layer include incorrect peripheral configuration or initialization, hardware-dependent mistakes, and timing-related problems such as an inappropriate sampling period for an external device. Timing-related bugs are particularly problematic because suspending execution with a debugger may itself perturb the timing behavior and reduce reproducibility. This suggests that the framework should support not only stop-and-inspect debugging but also forms of observation that do not rely on intrusive suspension.

\subsection{Requirements}

The observations in the previous subsection lead to several requirements for an Emfrp debugger. First, because the I/O layer and the intermediate layer are written in C, the framework should cooperate with an existing C debugger rather than replace it entirely (Req-1). Second, for state observation at the Emfrp level, the framework should display the current execution position in Emfrp source code (Req-2), show Emfrp variables such as nodes and local variables despite identifier transformations in generated code (Req-3), visualize the dependency graph to help developers understand state and state transitions (Req-4), provide stack traces at the level of Emfrp functions rather than C functions (Req-5), and support inspection of expression evaluation (Req-6). Third, for execution control, the framework should support multiple granularities of stepping, including stepping by subexpression, by node update, and by iteration (Req-7), as well as watchpoints for changes in node values (Req-8) and breakpoints at multiple granularities (Req-9). Finally, for bug reproduction, the framework should support input-trace-driven execution (Req-10). Together, these requirements motivate a multi-mode debugging framework that combines Emfrp-level debugging support with integration into existing C-level debugging infrastructure.

\section{Design and Implementation of Emdb}

\begin{figure*}
\centering\includegraphics[scale=0.85]{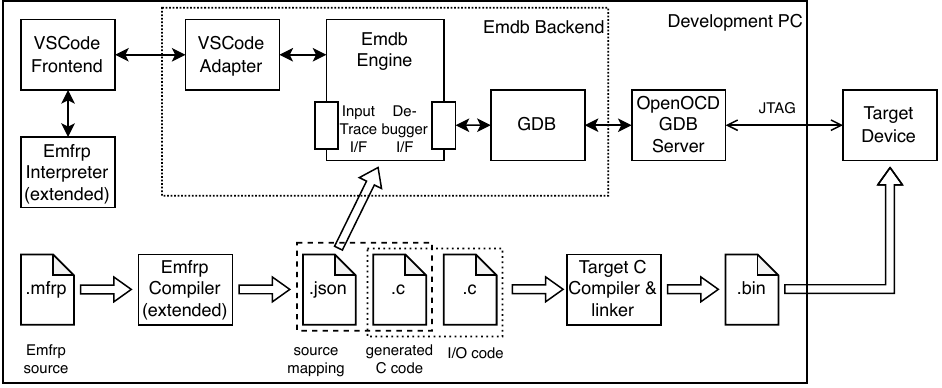}
\caption[Emdb Architecture]{Emdb Architecture}\label{fig:emdb}
\end{figure*}

\begin{figure*}
\centering\includegraphics[scale=0.4]{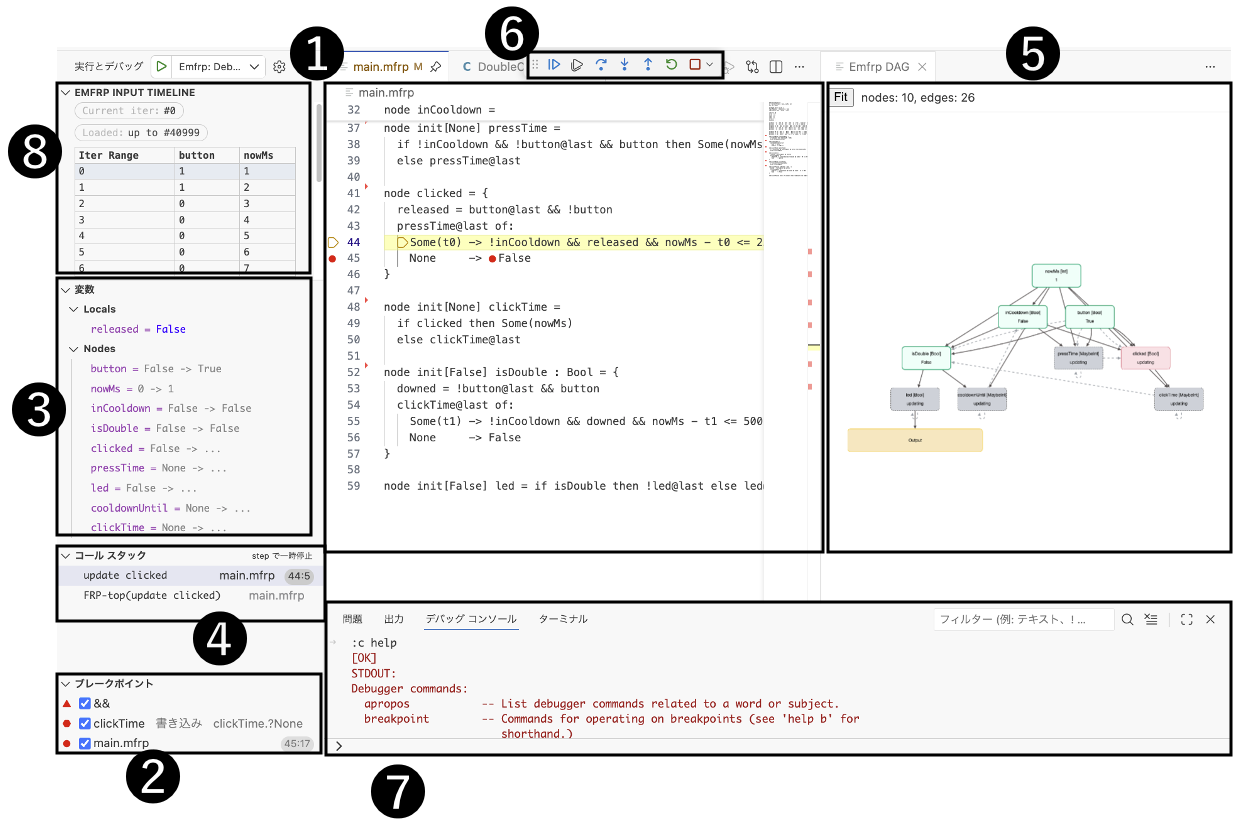}
\caption[VSCode Frontend of Emdb]{VSCode Frontend of Emdb:\;%
\ding{202} Emfrp source view,
\ding{203} breakpoint view,
\ding{204} node values view,
\ding{205} call stack view,\\
\ding{206} dependency graph view,
\ding{207} debug toolbar,
\ding{208} debug console, and
\ding{209} input-trace view
}\label{fig:vscodeview}
\end{figure*}

\subsection{Emdb Architecture}
Based on the requirements discussed in the previous section, we designed and implemented Emdb, a debugger for Emfrp programs. Figure~\ref{fig:emdb} shows the overall architecture of Emdb and the surrounding debugging environment. Emdb consists of two main components: a frontend provided as an extension for Visual Studio Code (VSCode), and a backend implemented in Python. The frontend accepts user operations such as continue, stop, step execution, and breakpoint management, and presents the execution state through multiple views (Figure~\ref{fig:vscodeview}). These views include the Emfrp source view, a breakpoint view, node values, a call stack, a dependency graph, a debug console, and an input-trace view. In particular, the source view displays the current execution position in Emfrp source code and supports breakpoints on source locations and subexpressions, while the dependency-graph view visualizes node states and updates them as execution proceeds. The debug console supports both Emfrp-level expression evaluation and low-level C debugger commands.

The backend interprets requests from VSCode and provides debugging functionality at the Emfrp level. Internally, it consists of a VSCode adapter, which translates requests from the frontend into backend API calls, and the Emdb engine, which implements the core debugging logic. The engine also provides an interface for registering input traces in advance when input-trace-driven execution is used. At its foundation, the Emdb engine relies on a debugger interface that controls an underlying C debugger.
In this sense, Emdb can be viewed as a wrapper that adds Emfrp-aware observation and control on top of a conventional C-level debugger. This design directly addresses Req-1: debugging of the I/O layer and the intermediate layer can be delegated to an existing C debugger, while generic mechanisms such as target control, stop-reason retrieval, and low-level expression evaluation can be reused. In our implementation, we provide standard GDB-interface backends based on GDB/MI and on the Python interface of LLDB, which makes Emdb reusable across multiple targets with little additional effort.

\subsection{Source Mapping and Emfrp-Level Debugging}
A key feature of Emdb is its support for source-level debugging of Emfrp programs.
To realize this feature, we extended the Emfrp compiler so that it generates not only C source code but also a JSON-based source mapping file. This mapping describes the correspondence between the generated C code and the original Emfrp source, and is the basis for reconstructing Emfrp-level execution from C-level debugger events. The mapping contains four kinds of information: (1) the Emfrp element corresponding to each relevant location in generated C code, (2) the correspondence between Emfrp variables and C variables, (3) the correspondence between Emfrp types and C-level representations, and (4) dependency information among nodes. In addition, we extended the Emfrp interpreter so that Emfrp expressions can be evaluated directly inside the debugger. Together, these extensions allow Emdb to present execution state, values, and types in terms of Emfrp rather than generated C.

A central concept in the source mapping is a \emph{checkpoint}. A checkpoint is a C-level execution location, identified at line granularity, that corresponds to a meaningful step in Emfrp-level execution. To establish this correspondence, we modified the Emfrp compiler so that each relevant evaluation step of an Emfrp expression is represented by a separate line in the generated C code. The compiler records these lines as checkpoints in the source mapping. Consequently, transitions between checkpoints in C execution order reproduce the corresponding evaluation order at the Emfrp level. The compiler also ensures that at most one checkpoint is assigned to each line of generated C code.

\begin{lstlisting}[language=Emfrp,style=small,float,caption={Emfrp Pattern-Matching Expression},label={code:patternmatching}]
v of:
  0 -> f(2, g(3))
  1 -> 4
  _ -> 0
\end{lstlisting}

\begin{lstlisting}[language=C,style=small,float,caption={Compiled C Code of Listing~\ref{code:patternmatching}},label={code:checkpoint}]
int _ret, _tmp001, _tmp002;
if (v == 0) {                  // CP
    _tmp001 = g_0(3);          // CP
    _tmp002 = f_0(2, _tmp001); // CP 
    _ret = _tmp002;
} else if (v == 1) {           // CP
    _ret = 4;                  // CP
} else {                       // CP
    _ret = 0;                  // CP
}
\end{lstlisting}

For example, consider the Emfrp pattern-matching expression shown in Listing~\ref{code:patternmatching}.
The compiler generates C code conceptually equivalent to Listing~\ref{code:checkpoint}.
Here, the lines marked CP are checkpoints. If \texttt{v} is zero, execution first reaches the checkpoint corresponding to the selection of the 0 branch, followed by the checkpoints corresponding to the evaluation of \texttt{g(3)} and then \texttt{f(2, g(3))}. Thus, although the nested function application in the original Emfrp expression is translated into multiple C statements using temporary variables, transitions between checkpoints preserve the evaluation steps and their order at the Emfrp level. This correspondence allows Emdb to implement subexpression-level stepping by controlling execution at the corresponding C-level checkpoints. Although this example uses only simple constant patterns, more complex patterns are decomposed during compilation, and checkpoints are generated for the evaluation steps corresponding to the decomposed parts of the pattern.

The same principle is applied at different granularities. At the FRP top level, checkpoints are placed on lines corresponding to input processing, output processing, node initialization, and node-update function calls. Transitions among these checkpoints therefore correspond to node-level stepping along the dependency order. Inside node-update functions, checkpoints are introduced for expression evaluation as illustrated above. More complex constructs, including conditional expressions and pattern matching, are compiled so that their relevant evaluation steps are similarly represented by checkpoints. Using the source mapping between Emfrp constructs and these checkpoints, Emdb observes C-level execution through the GDB interface and presents it as execution at the Emfrp source level. The correspondence between checkpoints in the generated C code and locations in the original Emfrp source code is recorded in a JSON source-mapping file generated separately by the Emfrp compiler.

\section{Case Study: Finding a Bug in \texttt{DoubleClick}}


Listing~\ref{code:doubleclick} contains a subtle bug, and in this section we illustrate how Emdb can be used to locate it. The target program, \texttt{DoubleClick}, controls an LED using a single button. A click is recognized only when the button is pressed for at most 250 ms, and a double click is recognized when the second click occurs within 500 ms. After a double click is detected, the system is intended to ignore further button inputs for a 500 ms cooldown period. However, the program in Listing~\ref{code:doubleclick} does not work as expected: if a third button press occurs within 500 ms after a double click, the LED is toggled again. We confirmed this behavior on the target device. This example is representative of timing-dependent bugs that are important in embedded-system development and difficult to diagnose with conventional low-level debugging alone.
The debugging procedures described in this section were actually performed using an ESP32 microcontroller board (ESP32 DevkitC).

We use Emdb to debug this behavior as follows. First, Emdb is configured to record an input trace, that is, a sequence of external inputs observed during execution. While this recording feature is enabled, we perform a triple click on the real device and confirm the faulty behavior. Next, we replay the program on the host PC using the recorded trace. This allows us to reproduce the timing-dependent bug deterministically without relying on repeated manual input on the device. To reach the point where the first double click is recognized, we set a conditional breakpoint on the event that node isDouble becomes true. Equivalently, a breakpoint can be placed on the then branch of the if expression in the definition of \texttt{led}. After execution stops there, we set another breakpoint, or watchpoint, on the expression \lstinline[language=Emfrp]|!btn@last && btn|, which represents the third button press. Continuing execution then brings the debugger to the third press event. Figure~\ref{fig:vscodeview} shows the resulting state. 

At this point, we inspect the execution while displaying the dependency-graph view and stepping over execution at the granularity of node updates. 
The graph view indicates that the node \texttt{isDouble} becomes \texttt{True} again at the third button press, which should not happen if the cooldown mechanism is working correctly. The same figure also shows the current values of nodes such as \texttt{nowMs} and \texttt{btn}, together with the nodes currently being updated, making it easier to understand the propagation of state through the dependency graph. This immediately reveals that the unexpected LED toggle is caused by the incorrect evaluation of \texttt{isDouble}.

To determine why \texttt{isDouble} becomes true, we restart execution using the same recorded input trace and stop again at the same point. Emdb supports deterministic re-execution from a previously saved input trace, so the same execution state can be reproduced without recreating the interaction manually. We then step into the definition of \texttt{isDouble} and examine the branch selected in the pattern match on \lstinline[language=Emfrp]|clickTime@last|. The debugger shows that execution reaches the branch corresponding to \texttt{Some(t)}, meaning that a previous click time is still stored. At this point, the local variable corresponding to the most recent press event indicates that the button has just been pressed, while the values of \texttt{t} and \texttt{nowMs} show that the current press still falls within 500 ms of the stored click time. As a result, the condition for \texttt{isDouble} evaluates to true even though the system should already be in the cooldown period.

This observation reveals the actual cause of the bug. The previous click time is preserved in \texttt{clickTime}, so the third press is still interpreted as occurring within 500 ms of the previous click. In other words, the cooldown mechanism prevents clicked from being updated, but it does not invalidate the previously stored click time. As a result, the old click timestamp remains available to isDouble, and the cooldown is effectively bypassed. To fix this problem, either isDouble must explicitly check \texttt{inCooldown}, or \texttt{clickTime} must be cleared when a double click is detected. In our implementation, we adopt the former solution by adding \lstinline|!inCooldown| to the \lstinline|Some(t)| branch of \texttt{isDouble}. After this modification, the program behaves correctly, which we confirmed both in replay-based debugging and on the target device. 

We discuss the qualitative benefits of Emdb compared with using GDB directly on the generated C code. When a conventional C debugger is used, understanding Emfrp-level state requires the developer to manually reconstruct correspondences between Emfrp nodes or local variables and the identifiers and values that appear in the generated C code. In contrast, Emdb directly presents node values and local variables at the Emfrp level through its variable view. Similarly, C-level debugging requires the user to manually translate Emfrp expressions, identifiers, and types when setting breakpoints, conditional breakpoints, and stepping through execution. Emdb eliminates this burden by providing these operations directly at the Emfrp level. In addition, Emdb supports input-trace-driven execution, which allows developers to replay and modify input sequences without regenerating logs, editing C code, or rebuilding the program. These features substantially reduce the amount of auxiliary work required during debugging.

\section{Related Work}

Several existing debugging approaches are related to Emdb, but they target different execution models and debugging granularities. Salvaneschi and Mezini proposed Reactive Inspector~\cite{Salvaneschi:2016aa}, a debugger for the Scala FRP library REScala~\cite{Salvaneschi:2014ab}. Like Emdb, Reactive Inspector emphasizes visualization of dependency graphs and navigation over them as primary debugging operations. However, the transition model differs from that of Emdb. In REScala, recomputation is triggered only when the value of a dependent signal changes, so graph transitions correspond to propagation caused by value changes. In contrast, Emfrp reevaluates nodes in every iteration regardless of whether dependent values change, making iteration-based traversal of the dependency graph more natural. Moreover, Reactive Inspector does not explicitly provide observation of evaluation inside node definitions, whereas Emdb supports debugging at subexpression granularity.

Perez and Nilsson proposed a debugger for Yampa~\cite{Perez:2017aa}, an FRP library in Haskell. Their approach exploits the property that replaying the same discrete sequence of input events yields the same execution result, allowing past executions to be reproduced by saving and reusing input traces. They also combine this idea with QuickCheck to automatically generate event sequences and search for counterexamples. This replay-based view of debugging is closely related to Emdb's input-trace-driven execution, which likewise supports reproducible debugging of reactive behavior.

Rojas Castillo et al. proposed a language-independent debugging approach based on a debugging-oriented control-flow graph (CFG) for WebAssembly~\cite{Rojas-Castillo:2025aa}. This work is closely related to ours because it also considers WARDuino~\cite{Singh:2019aa}, a WebAssembly runtime for microcontrollers, as a target platform and thus shares our interest in debugging under the constraints of embedded systems. However, although their approach supports multiple source languages, it is primarily aimed at procedural languages compiled to WebAssembly. Emdb, in contrast, targets a pure FRP language with a substantially different execution model, requiring specialized source mappings, stepping semantics, and visualizations tailored to FRP computation.

Iyenghar et al. proposed model-based design-level debugging for embedded systems developed in a model-driven style~\cite{Iyenghar:2017aa}. Their approach consists of a lightweight target-side monitor and a host-side GUI debugger that visualizes target behavior at the design-model level using runtime traces. The design philosophy is similar to one mode of Emdb, in which execution observed on the target can be reconstructed and inspected on the host. However, their approach assumes UML-based design models and focuses on RTOS-level events, whereas Emdb reconstructs execution at the level of an FRP language. Their discussion of communication interfaces is also relevant: they compare serial communication and JTAG, and report that JTAG provides lower overhead, while buffering strategies can further reduce cost for high-frequency events. These observations are useful when considering communication mechanisms for Emdb.

\section{Conclusion and Future Work}

This paper presented a multi-mode debugging framework for FRP-based embedded systems and its implementation, Emdb, for the Emfrp language. Our approach addresses the abstraction gap that arises when Emfrp programs are compiled into C and combined with platform-specific C/C++ I/O code, forcing developers to debug a mixed low-level implementation rather than the original FRP program. By integrating Emfrp-level observation and control with an existing C debugger, Emdb enables developers to inspect execution in terms of nodes, dependencies, and temporal behavior while still retaining access to the underlying implementation when necessary.

The case study demonstrates that Emdb can effectively support the diagnosis of timing-dependent bugs that are difficult to locate with conventional C-level debugging alone. Our preliminary evaluation also suggests that Emdb substantially improves the practicality of debugging Emfrp programs, while keeping the overhead of several debugging operations sufficiently low for practical use.

There are several directions for future work. First, we plan to conduct a more systematic quantitative evaluation, including controlled comparisons of debugging efficiency. Second, although our current experiments mainly target Espressif ESP32 platforms, we would like to validate the approach on a wider range of embedded targets. Finally, we plan to investigate more unified source-mapping techniques for cross-paradigm debugging, rather than constructing language-specific mappings individually, for example by exploring whether approaches such as CFG-based representations can be extended to FRP and other programming paradigms.

We have also developed several extensions to Emfrp, including language constructs that facilitate the description of periodic and aperiodic tasks, which frequently arise in embedded systems~\cite{Sogo:2023aa}. As discussed in the case study, Emdb can already support debugging of timing-sensitive behavior by recording and replaying input traces. However, explicitly supporting such timing-oriented language extensions poses additional challenges, for example in presenting task activation, timing relationships, and execution progress at the Emfrp source level. Extending Emdb to provide effective debugging support for these constructs is therefore an important direction for future work.

\begin{acks}
The authors would like to thank the anonymous reviewers for their valuable comments.
We are also grateful to Go Suzuki for his suggestions regarding microcontroller-based systems.
This work was supported in part by JSPS KAKENHI Grant Numbers JP22K11967 and JP24K14892.
\end{acks}

\bibliographystyle{ACM-Reference-Format}
\bibliography{refs}


\end{document}